\documentclass[12pt]{article}
\usepackage{array}
\usepackage{graphicx}
\usepackage{amssymb}
\usepackage{amsmath}
\usepackage{multirow}
\usepackage{cite}
\def\@fmsl@sh#1#2#3{\m@th\ooalign{$\hfil#1\mkern#2/\hfil$\crcr$#1#3$}}
 \def\eq#1\en{\begin{equation}#1\end{equation}}
\def\s[#1,#2]{[#1\stackrel{\star}{,}#2]}
\def\sx[#1,#2]{[#1\stackrel{\star_{x}}{,}#2]}

\newcommand{\nc}{\newcommand}
\nc{\beq}{\begin{equation}}
\nc{\eeq}{\end{equation}}
\nc{\beqa}{\begin{eqnarray}}
\nc{\eeqa}{\end{eqnarray}}

\def\bc{\begin{center}}
\def\ec{\end{center}}

\def\gsim{\mathrel{\mathpalette\atversim>}}

\def\bc{\begin{center}}
\def\ec{\end{center}}

\def\gsim{\mathrel{\rlap{\lower4pt\hbox{\hskip1pt$\sim$}}

    \raise1pt\hbox{$>$}}}       

\def\gsim{\mathrel{\rlap{\lower4pt\hbox{\hskip1pt$\sim$}}
    \raise1pt\hbox{$>$}}}       

\begin{document}
\makeatletter
\def\fmslash{\@ifnextchar[{\fmsl@sh}{\fmsl@sh[0mu]}}
\def\fmsl@sh[#1]#2{%
  \mathchoice
    {\@fmsl@sh\displaystyle{#1}{#2}}%
    {\@fmsl@sh\textstyle{#1}{#2}}%
    {\@fmsl@sh\scriptstyle{#1}{#2}}%
    {\@fmsl@sh\scriptscriptstyle{#1}{#2}}}
\def\@fmsl@sh#1#2#3{\m@th\ooalign{$\hfil#1\mkern#2/\hfil$\crcr$#1#3$}}
\makeatother

\thispagestyle{empty}
\begin{titlepage}
\boldmath
\begin{center}
  \Large {\bf Primordial Black Holes and a Large Hidden Sector}
    \end{center}
\unboldmath
\vspace{0.2cm}
\begin{center}
{ 
{\large Xavier Calmet}\footnote{x.calmet@sussex.ac.uk}
}
 \end{center}
\begin{center}
{\sl Physics and Astronomy, 
University of Sussex,  \\ Falmer, Brighton, BN1 9QH, UK 
}
\end{center}

\vspace{\fill}
\begin{abstract}
\noindent
In this note we point out that primordial black holes could be much shorter lived than usually assumed if there is a large hidden sector of particles that only interacts gravitationally with the particles of the standard model. The observation of the explosion of one of these black holes would severely constrain the energy scale at which gravity becomes strong.
\end{abstract}  
{\bf keywords:} primordial black holes, particle physics models.

\end{titlepage}



\newpage

Primordial black holes have been proposed as a means to probe fundamental physics using astronomical observations \cite{Carr:1975qj,Carr:2003bj,Khlopov:2008qy}. They also have been considered as potential dark matter candidates \cite{Blais:2002nd}.
In this note we point out that primordial black holes could be much shorter lived than usually assumed if there is a large hidden sector of particles that only interact gravitationally with the particles of the standard model.  There are two important physical quantities which can affect the lifetime of black holes, given a large hidden sector. The mass spectrum of the particles in the large hidden sector and the temperature of the large hidden sector determine if black holes can decay in that sector via Hawking radiation.
A much shorter lifetime is expected if some of the masses of the hidden sector particles are smaller than the temperature of the black hole when it is created and if the temperature of the hidden sector is below that of the black hole. 

It is now well accepted that a black hole can decay through the emission of Hawking radiation \cite{Hawking:1974sw}.  The time it takes for a black hole
to decay is given in the standard model of particle physics by (see e.g. \cite{Wald:1984rg})
\begin{eqnarray}
t=\frac{5120 \pi G^2 M_{BH}^3}{\hbar c^4}
\end{eqnarray}
where $G$ is Newton's constant and $M_{BH}$ is the black hole mass.  As we shall see shortly, the lifetime of  a black hole can be affected if the available phase-space is larger than in the standard model. A black hole of one solar mass (i.e $M_{BH} \approx 2 \times 10^{30} \ \mbox{kg}$ will decay in $2.1 \times 10^{67} \ \mbox{years}$, which is much larger than the age of the universe ($13.7 \times 10^9 \ \mbox{years}$). However, a black hole today can only evaporate if its temperature is above the temperature of the cosmic microwave background today i.e. $2.7 \ \mbox{K}$. The temperature of a black hole is related to its mass by
\begin{eqnarray}
T_H=\frac{\hbar c^3}{8 \pi G  M_{BH} k_B}.
\end{eqnarray}
 This implies that only black holes with masses less than $0.8 \%$ of the earth mass can decay today. A black hole of $10^{11} \mbox{kg}$ would evaporate in 2.7 billion years. Exploding primordial black holes could thus be discovered. However, if there is a large hidden sector of particles in nature that only interacts gravitationally with those of standard model, black holes could evaporate much faster.  If the phase-space for the black hole to decay is increased by a factor $N$, where $N$ stands for the number of new particles, one has approximatively 
\begin{eqnarray}
t\approx \frac{5120 \pi G^2 M_{BH}^3}{\hbar c^4} \frac{1}{N}.
\end{eqnarray}
If $N$ is large, i.e.,  if there are some $10^9$ and some of the masses of the particles of the hidden sector are below the temperature of the black hole,  primordial black holes would have decayed a long time ago and an observation of a primordial black hole exploding today would imply a tight bound on the number of fields in a hidden sector. Note that as the black hole decays the phase-space will open up to more massive particles in the hidden sector as the temperature of the black hole will increase as its mass decreases.

 This is of particular interest for models of particles physics which are addressing the gauge hierarchy problem by introducing a large number of fields. This mechanism effectively reduces the energy scale of quantum gravity to 1 TeV, see \cite{Calmet:2010nt} for a review. This requires some $10^{33}$ new fields with masses below 1 TeV. Note that the large hidden sector does not necessarily need to be in thermal equilibrium with the visible universe since its particles only interact gravitationally with the standard model fields and its temperature could be lower or higher than the CMB temperature. In any case, if there is a large hidden sector, primordial black holes will decay invisibly from the standard model point of view. If the temperature of the hidden sector is closer to absolute zero, more massive black holes could decay via Hawking radiation in that sector, assuming that the mass spectrum of the hidden sector allows it.

An observation of an exploding primordial black hole today would allow us to constrain the scale at which quantum gravity becomes strong which is defined by $\bar M(\mu_\star)\sim \mu_\star$ where $\bar M(\mu_\star)$ is the running reduced Planck mass given by \cite{Calmet:2008tn,Atkins:2010eq}
\begin{eqnarray} 
\bar M(\mu)^2= \bar M(0)^2-\frac{\mu^2}{96 \pi^2} N_l 
\end{eqnarray}
where we set $\hbar=c=1$, $\bar M(0)$ is the reduced Planck mass  measured over astrophysical distances and $N_l=N_S+N_F-4 N_V$ with $N_S$, $N_F$ and $N_V$ respectively the number of real scalars, Weyl fermions and vector bosons in the theory. The true energy scale  $\mu_*$ at which quantum gravity effects are large is one at which 
\begin{equation}
\bar M^2 _P(\mu_*) \sim \mu_*^{2}.
\end{equation} 
This condition implies that fluctuations in spacetime geometry at length scales $\mu_*^{-1}$ will be unsuppressed. 
 One finds \cite{Calmet:2008tn,Atkins:2010eq,Calmet:2008df}:
\begin{eqnarray}
\mu_\star= \frac{\bar M(0)}{\sqrt{1+ \frac{N_l}{96 \pi^2}}}.
\end{eqnarray}
This shows that the energy scale at which quantum gravitational effects become large, i.e. the Planck scale, can be much lower than naively assumed. In an extreme case, the Planck scale physics could be relevant for  the LHC. As in models with large extra-dimensions, the Planck mass could be in the TeV region. If there is a large hidden sector consisting of $10^{33}$ particles of spin 0 and/or 1/2 which are only interacting gravitationally with the standard model, the scale of quantum gravity  is at 1 TeV and quantum black holes \cite{Calmet:2008dg} or gravitons \cite{Calmet:2009gn} could be produced at the LHC. Newton's potential has been probed up to distances of $(10^{-3} \mbox{eV})^{-1}$, the masses of the bulk of the particles of the hidden sector should be larger than $10^{-3}$ eV otherwise the running of Newton's constant would have been observed in e.g. the E\"ot-Wash short-range experiment. Note that cosmic ray experiments  set a bound on the four-dimensional Planck mass of the order of 500 GeV \cite{Calmet:2008rv} since small black holes could form in the collision of cosmic rays with nuclei in the atmosphere.

The temperature of a $0.8 \%$ earth mass black hole is 0.1 GeV.  If a decaying  primordial black hole was observed, we would conclude that $N_l\sim N<10^9$ with at least some of the masses below $\sim 0.05 \ \mbox{GeV}$ and thus $\mu_\star>2.4 \times 10^{15} \ \mbox{GeV}$. We provide some more numerical examples in Table 1.  The mass of a primordial black hole in $kg$ created in the early universe and decaying today as a function of the number of fields $N$ in the hidden sector is shown in Figure 1.  Note that the case of $N\sim 10^{33}$ particles of masses below $10^{-20} \ \mbox{GeV}$ is ruled out by probes of Newton's 1/r potential in the solar system. As the black hole decays, its temperature will increase thus opening decay channels into heavier particles. The extreme case would be that of a quantum black hole with a mass of the order of $\mu_\star$ which could decay into two particles of masses $\mu_\star/2$. It is thus not strictly speaking necessary that all particles of the hidden sector are lighter than the original temperature of the black hole as long as some of them are in order to allow the temperature to increase. Generically speaking, black holes which are cooler than the CMB will evaporate into the hidden sector if the mass spectrum of the hidden sector particles allows for it.

 \begin{table*}[tbh]
\resizebox{\textwidth}{!}{
\begin{tabular}{|c|c|c|c||c|}  
\cline{5-5}
\multicolumn{1}{r}{} & \multicolumn{1}{r}{}& \multicolumn{1}{r}{}&\multicolumn{1}{r}{} & \multicolumn{1}{|c|}{reduced Plank mass } \\
\hline
BH mass in kg    & $5 \times 10^{6}$& $10^{11}$&$10^{30}$ & \\
 \hline
  Temperature in GeV & $2 \times10^3$& 0.1 & $10^{-20}$    & \\
  \hline
     lifetime in $10^9$ yr, $N=0$ &$10^{-13}$ &2.7 & $ 10^{57}$ &  $2.43 \times 10^{18}$ GeV\\  
\hline
  lifetime in $10^9$ yr, $N={10^9}$ &$10^{-22}$ &$10^{-9}$ & $10^{48}$& $2.4 \times 10^{15}$ GeV   \\  
\hline
lifetime in $10^9$ yr, $N={5\times 10^{33}}$ &$10^{-47}$ &$10^{-34}$ & $ 10^{23}$ & 1000 GeV  \\  
\hline
\end{tabular}}
\caption{$N$ is the number of particles with masses below  the final black hole temperature.}  \label{t1}
\end{table*}

 \begin{figure}
\center
\includegraphics[scale=1]{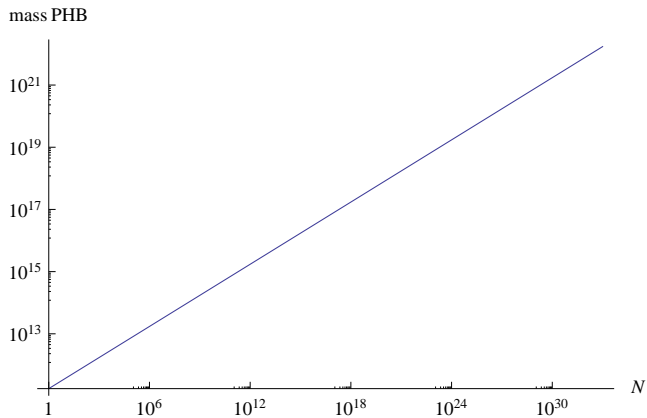}
\caption{This double logarithmic scale graph shows the mass of a primordial black hole in kg created in the early universe and decaying today as a function of the number of fields $N$ in the hidden sector.}
\label{pic1}
\end{figure}

Clearly an observation of an exploding primordial black hole of mass $\sim 10^{11}$ kg would rule out a hidden sector of more than $10^9$ particles with  a mass spectrum starting below $0.05$ GeV. As the black hole decays, more massive particles would become accessible. We would conclude that the  Planck mass has to be larger than $2.4 \times 10^{15}$ GeV. Note, however, that it is possible to build a model with $10^9$ particles and masses above $0.1$ GeV or with a temperature of the hidden sector higher than the CMB temperature in which case the lifetime of primordial black holes would be less drastically affected as the hidden sector particles only become relevant when the black hole mass has already sufficiently decreased. In any case, the observation of a decaying primordial black hole today would severely constrain the parameter space of a large hidden sector. This represents yet another strong motivation to search for decaying primordial black holes. 

Finally let us conclude by pointing out that if the temperature of the hidden sector is very close to absolute zero, any black hole can decay today into the hidden sector via Hawking radiation assuming that some of the particles in the hidden sector are light enough. The lifetime of astrophysical black holes could be sizably affected by a large hidden sector. Similar observations have been made in the framework of  extra-dimensional models \cite{Emparan:2002jp} which effectively also contain a large number of fields.

\bigskip

{\it Acknowledgments:} 
 It is my pleasure to acknowledge discussions with Stephen Hsu and Claus Kiefer.
This work is supported in part by the European Cooperation in Science and Technology (COST) action MP0905 ``Black Holes in a Violent Universe". 



\bigskip


\end{document}